\documentstyle[preprint,aps]{revtex}
\def\slash#1{\ooalign{\hfil/\hfil\crcr$#1$}}
\begin{document}
\preprint{\vbox {\hbox{RIKEN-AF-NP-295}}}

\title{Nonperturbative QCD corrections to the effective coefficients of the four-Fermi 
operators} 
\author{Mohammad R. Ahmady$^a$\footnote{Email: ahmady@riken.go.jp}and Victor Elias$^b$\footnote{Email:zohar@apmaths.uwo.ca}}
\address{
$^a$LINAC Laboratory, The Institute of Physical
and Chemical Research (RIKEN)\\ 2-1 Hirosawa, Wako, Saitama 351-0106,
Japan \\
$^b$Department of Applied Mathematics, University of Western Ontario\\
London, Ontario, Canada N6A 5B7}

\date{January 1999}
\maketitle
\begin{abstract}
We calculate the leading nonperturbative contributions to the effective Wilson 
coefficient of the four-fermion operators arising from the QCD penguin, and we demonstrate
how the usual perturbative one loop contribution is augmented by nonperturbative
condensates.  These corrections, 
which are obtained by quark and gluon condensate insertions into the quark loops, 
enter at the next-to-leading logarithm precision.   
Our results indicate for the charmed quark loop that the gluon condensate contribution is
quite sensitive to the momentum transfer to the quark-antiquark pair.   
\end{abstract}
%
\newpage
Nonleptonic B decays to light hadrons have recently received a lot of theoretical
attention.  This is mostly due to the measurements by the CLEO Collaboration of a number 
of $B\to h_1h_2$ decays, where $h_1$ and $h_2$ are light hadrons such as $\pi$, $K$, $\eta'$ 
and $\omega$\cite{s,behrens,br}.  For example, the larger than expected experimental results 
for the branching ratio of the inclusive $B\to X_s\eta'$ and its exclusive counterpart $B\to
K\eta'$ have led to various theoretical explanations both within and beyond the Standard
Model\cite{as,ht,hz,yc,aks}.  On the other hand, these decays could provide the first
evidence for CP violation in B decays in the near future\cite{ak}.

The standard theoretical framework to study the nonleptonic $B$ decays is based on the
effective Hamiltonian approach.  In this way, the heavy degrees of freedom, i.e. top quark 
and $W^\pm$ gauge bosons, are integrated out and the matrix element can be written as
\begin{equation}
<h_1h_2\vert H_{\rm eff}\vert B>=\frac{G_F}{\sqrt{2}}\sum_iV^i_{\rm CKM}C_i(\mu ) <h_1h_2\vert O_i(\mu )\vert B> 
\end{equation}
where $\{ O_i\}$ is the complete set of operators relevant to $B\to h_1h_2$ decay and 
$C_i(\mu)$ are the corresponding Wilson coefficients evaluated at the renormalization 
scale $\mu$\cite{buras}.  These coefficients have been calculated up to the next-to-leading
logarithm (NLL) precision in the strong coupling constant $\alpha_s$ \cite{b,cfmr}.  
At this precision, the matrix elements on the RHS of that equation should be evaluated 
at the one loop level in order to have a renormalization-scheme independent result for the 
LHS of the eq. (1) \cite{b,kps}.

In this work, we concentrate on the four-Fermi operator subset of $\{ O_i\}$, i.e. $i=1..6$, 
which is relevant for the hadronic $B$ decays.  The first two operators $O_1$ and $O_2$ 
are the current-current opertors:
\begin{equation}
O_1=\bar q_\alpha\gamma^\mu Lu_\alpha\bar u_\beta\gamma_\mu Lb_\beta \; ,\; 
O_2=\bar q_\alpha\gamma^\mu Lu_\beta\bar u_\beta\gamma_\mu Lb_\alpha \;\; ,
\end{equation}
and the other four are the so-called QCD penguin operators:
\begin{eqnarray}
\nonumber O_3&=&\bar q_\alpha\gamma^\mu Lb_\alpha\sum_{q'}\bar q_\beta'\gamma_\mu Lq_\beta' 
\; ,\; O_4=\bar q_\alpha\gamma^\mu Lb_\beta\sum_{q'}\bar q_\beta'\gamma_\mu Lq_\alpha' \;\; ,
\\ O_5&=&\bar q_\alpha\gamma^\mu Lb_\alpha\sum_{q'}\bar q_\beta'\gamma_\mu Rq_\beta '\; ,\; 
O_6=\bar q_\alpha\gamma^\mu Lb_\beta\sum_{q'}\bar q_\beta'\gamma_\mu Rq_\alpha' \;\;.
\end{eqnarray}
where $q=s$ or $q=d$ for $b\to s$ or $b\to d$ transitions, respectively, and where $L$ and $R$ 
are the left and right handed projection operators, i.e. $L=1-\gamma_5$ and $R=1+\gamma_5$.  Two more current-current operators which are obtained by replacing $u$ quark by $c$ quark in eq. (2) should also be added to the above list.  

In NLL precision, the matrix  element of the operators is to be 
evaluated at one loop level, as mentioned above.  These one loop matrix elements 
in turn can be written in terms of tree level matrix elements with loop effects 
included in the corresponding effective Wilson coefficients $C_i^{\rm eff}$\cite{b,kps}.  
There are two types of relevant one loop diagrams, the vertex corrections, where a gluon 
connects two of the outgoing quark lines, and penguin diagrams where a quark-antiquark 
pair forms a loop and emits a gluon which in turn produces a quark-antiquark pair.  
Our focus in this paper is on the latter type of corrections (Fig. 1) since they could have 
a potentially significant effect on branching ratios and CP asymmetry in two body hadronic 
$B$ decays\cite{kps,akl}.  In fact, the absorptive portion of penguin diagrams can provide 
the strong phase necessary  for direct CP violation.  This one loop correction is analogous 
to the so-called continuum contribution in the dileptonic rare $B$ decays 
$B\to X_s(X_d)\ell^+\ell^-$, where one has to deal with a lepton-antilepton pair, as opposed to a 
quark-antiquark pair \cite{ahm96}.  For the dilepton case, we also have significant nonperturbative 
contributions from intermediate vector mesons which are quite sensitive to the 
momentum-transfer to the dileptons.  Motivated by the above analogy, we embark on supplementing 
$C_i^{\rm eff}$ with a corresponding set of nonperturbative contributions.

In this work, we calculate the leading nonperturbative corrections to the perturbative evaluation of the penguin diagrams.  Ignoring the vertex corrections, only the QCD penguin operators receive corrections from one loop diagrams of Fig. 1\cite{kps}:
\begin{eqnarray}
\nonumber
C_3^{\rm eff}&=&C_3-\frac{1}{6}\frac{\alpha_s}{4\pi}\left (C_t+C_p \right ) +\; ...\;\; , \\
\nonumber
C_4^{\rm eff}&=&C_4+\frac{1}{2}\frac{\alpha_s}{4\pi}\left (C_t+C_p \right ) +\; ...\;\; , \\
\nonumber
C_5^{\rm eff}&=&C_5-\frac{1}{6}\frac{\alpha_s}{4\pi}\left (C_t+C_p \right ) +\; ...\;\; , \\
C_6^{\rm eff}&=&C_6+\frac{1}{2}\frac{\alpha_s}{4\pi}\left (C_t+C_p \right ) +\; ...\;\; .
\end{eqnarray}
$C_t$ and $C_p$ are the loop corrections corresponding to the current-current (tree) and 
penguin type four-fermion operator insertion in Figure 1, respectively.  Using naive 
dimensional regularization(NDR) along with the $\overline{\rm MS}$ renormalization scheme,
one obtains
\begin{eqnarray}
\nonumber
C_t&=& -\sum_{q'=u,c}\frac{V_{q'b}V_{q'q}^*}{V_{tb}V_{tq}^*}\left [\frac{2}{3}+\frac{2}{3}Ln(\frac{m^2_{q'}}{\mu^2})-\Delta F_1(\frac{k^2}{m^2_{q'}})\right ]C_1\;\; , \\
\nonumber
C_p&=& \left [\frac{4}{3}+\frac{2}{3}Ln(\frac{m^2_{s}}{\mu^2})+\frac{2}{3}Ln(\frac{m^2_{b}}{\mu^2})-\Delta F_1(\frac{k^2}{m^2_{s}})-\Delta F_1(\frac{k^2}{m^2_{b}})\right ]C_3 \\
&{}&+\sum_{q'=u,d,s,c,b}\left [\frac{2}{3}Ln(\frac{m^2_{q'}}{\mu^2})-\Delta F_1(\frac{k^2}{m^2_{q'}})\right ](C_4+C_6) \;\; ,
\end{eqnarray}
where the function $\Delta F_1(z)$ is defined as
\begin{equation}
\Delta F_1(z)=-4\int^1_0dxx(1-x)ln\left [1-zx(1-x)\right ]\;\; .
\end{equation}
We note that the effective coefficients in NLL depend on the momentum-transfer $k^2$ as 
well as the internal quark mass $m_q$. In fact, for $z\ge 4$, $\Delta F_1(z)$ develops 
an imaginary part which has been taken as the strong phase necessary for generating direct
CP asymmetry\cite{kps}.  Note that the light internal quarks produce the dominant 
contributions; consequently, the larger factor $C_t$ ($C_1$ is much larger then $C_{3...6}$) 
is more significant for $b\to d$ transitions in which the ratio of CKM factors for tree and 
penguin is not too small.  Nevertheless, one can not have a complete picture of 
the NLL corrections without taking into account nonperturbative effects which, as in 
the case of the intermediate resonance contribution to $B\to X_s(X_d)\ell^+\ell^-$, can 
prove to be significant.  To investigate this possibility, we calculate the leading nonperturbative 
contributions to $C_i^{\rm eff}\; ,\; i=3..6$ by inserting quark and gluon condensates, 
which are believed to be responsible for the chiral-symmetry breaking and confinement properties, 
respectively, of the QCD vacuum,  into one-loop penguin diagrams (Figs. 2,3). 

QCD condensates characterize the nonperturbative content of the QCD vacuum, and 
therefore, condensate insertions should give us a hint of the leading nonperturbative 
contributions to the effective Wilson coefficients.  To demonstrate our method, we 
present a detailed calculation of the quark condensate result.  For the gluon condensate 
contribution, the necessary expressions are extracted from previously determined \cite{blp}
condensate contributions to the vector correlation function. 

To obtain the quark condensate contribution, we proceed by replacing the usual 
perturbative internal quark propagator in Fig. 1 with the full quark propagator 
$S(p)$:
\begin{equation}
S(p)=S^{\rm P}(p)+S^{\rm NP}(p).
\end{equation}
The nonperturbative contributions to the quark propagator can then be written as
\begin{equation}
iS^{\rm NP}(p)=\int d^4xe^{ip.x}<\Omega\vert :\Psi (x)\bar\Psi (0):\vert\Omega >\;\; ,
\end{equation}
where $\vert\Omega >$ is the nonperturbative QCD vacuum. 
The nonlocal vacuum expectation value (vev) in eq. (8) can be expanded in terms of 
local condensates\cite{ess}.  To ascertain how the leading nonperturbative QCD 
contributions affect the coefficients $C_i^{\rm eff}$, only the lowest dimensional 
(quark and gluon condensates) components of the expansion are considered below.  
However, we note that the nonlocal vev of two quark fields does not contain a gluon 
condensate component.  The quark condensate projection of the nonperturbative quark 
propagator is taken from Ref. \cite{baes}:
\begin{equation}
iS^{\rm NP}(p)={(2\pi )}^4(\slash{p} +m_q)F(p)
\end{equation}
where the Fourier transform of $F(p)$ is expressed in terms of Bessel function
\begin{equation}
\int d^4pe^{-ip.x}F(p)=-\frac{<\bar qq>}{6m_q^2}\frac{J_1(m_q\sqrt{x^2})}{\sqrt{x^2}}\;\; ,
\end{equation}
with the following important property:
\begin{equation}
(p^2-m_q^2)F(p)=0\;\; .
\end{equation}
The dimension-3 quark condensate $<\bar qq>$ is the vev of the normal ordered local 
product of quark and antiquark fields, i.e.
\begin{equation}
<\bar qq>=<\Omega\vert :\bar\Psi (0)\Psi (0):\vert\Omega >\;\; .
\end{equation}
Using the Feynman rule of eq. (9), one can obtain the nonperturbative $<\bar qq>$ 
contribution to the effective Wilson coefficients in a straightforward manner.  
The relevant Feynman diagram is illustrated in Fig. 2, where the nonperturbative 
quark propagator $S^{\rm NP}$ is depicted by a disconnected line with two dots.   
Here we concentrate on the loop portion of Fig. 2 which differs from the 
perturbative case of Fig. 1.  Aside from the color factor, the vector current correlation 
function can be written as
\begin{equation}
\Pi_{\mu\nu}^{<\bar q q>}(k)=2\int d^4p\frac{Tr[(\slash{p}-\slash{k}+m)\gamma_\mu (\slash{p}+m)\gamma_\nu ]F(p)}{{(p-k)}^2-m^2+i\epsilon}\;\; ,
\end{equation}
where the factor 2 in front is due to two possible insertions of the nonperturbative 
quark propagator in the fermion loop.  By contracting 
$p^\mu p^\nu$ into $\Pi_{\mu\nu}^{<\bar qq>}$ and using the identity
\begin{eqnarray}
\nonumber
\int d^4pk.pF(p)&=&i\lim_{\xi\to 0}\frac{d}{d\xi}\int d^4pe^{-i\xi k.p} F(p)\\
&=&-i\frac{<\bar qq>}{6m_q}\lim_{\xi\to 0}\frac{d}{d\xi}\frac{J_1(m_q\xi\sqrt{k^2})}{m_q\xi\sqrt{k^2}}=0\;\; ,
\end{eqnarray}
where the second line is derived via eq. (10), one can show explicitly the 
transversality of the correlation function in eq. (13).  Consequently, one finds upon
contracting $g^{\mu\nu}$ into $\Pi_{\mu\nu}^{<\bar qq>}$ and imposing the on-shell 
constraint (11) that
\begin{eqnarray}
{\Pi^{<\bar qq>}}_\mu^\mu (k) =24(2m_q^2+k^2)\int d^4p\frac{F(p)}{k^2-2k.p+i\epsilon}-24\int d^4pF(p)\;\; .
\end{eqnarray}
The integrals appearing in eq. (15) are evaluated as follows\cite{baes}:
\begin{eqnarray}
\nonumber
\int d^4pF(p)&=&-\frac{<\bar qq>}{12m_q}\;\; , \\
\nonumber
\int d^4p\frac{F(p)}{k^2-2k.p+i\epsilon}&=&\frac{i<\bar qq>}{12m_q^2\sqrt{k^2}}\int^\infty_0\frac{d\eta}{\eta}e^{i\eta k^2-\epsilon\eta}J_1(2\eta m_q\sqrt{k^2}) \\
&=&-\frac{<\bar qq>}{6m_qk^2}{\left [1+\sqrt{1-\frac{4m_q^2}{k^2}}\right ]}^{-1}\;\; ,
\end{eqnarray}
where the final line is derived by utilizing a tabulated integral\cite{gr}.  
Using the transversality of $\Pi_{\mu\nu}^{<\bar qq>}$, the result for the RHS of (13)
is easily obtained:
\begin{equation}
\Pi_{\mu\nu}^{<\bar qq>}(k)=-\frac{<\bar qq>}{3m_q^3}\left [1-\left (1+\frac{2m_q^2}{k^2}\right )\sqrt{1-\frac{4m_q^2}{k^2}}\right ](g_{\mu\nu}k^2-k_\mu k_\nu )\;\; ,
\end{equation}
This result is, in fact the $<\bar{q}q>$-contribution to the vector-current correlation function derived in
ref. \cite{blp}.  The corresponding gluon condensate contribution to the vector-current correlator,
which may be derived by several different methods \cite{baes}, is given by \cite{blp}
\begin{eqnarray}
\Pi_{\mu\nu}^{<G^2>} (k) & = & (k_\mu k_\nu / k^2 - g_{\mu\nu}) \frac{\alpha_s <G^2>}{12 \pi k^2 (1 - 4/z)^2} \nonumber\\
& \times & \left\{ \frac{12}{z} \left( \frac{1}{z^2} - \frac{2}{z^4} \right) X (z)
- \left(1 - \frac{4}{z} + \frac{12}{z^2} \right) \right\}
\end{eqnarray}
where
\begin{equation}
z \equiv k^2/m_q^2
\end{equation}
and where \cite{blp,asd}
\begin{eqnarray}
X(z) & \equiv & \frac{1}{(1-4/z)} \left[ \int_0^1 dx \;  ln [1 - z x (1-x) - i | \epsilon | ] + 2 \right] \nonumber\\
& = & \left[
\begin{array}{ll} \frac{1}{\sqrt{1+4/|z|}} ln \left[ \frac{\sqrt{1+4/|z|}\;+1}{\sqrt{1+4/|z|}\; -1} \right], & z < 0 \\
\frac{2\sqrt{4/z-1}}{(1-4/z)} tan^{-1} \left[ \frac{1}{\sqrt{4/z-1}} \right], & 0 < z < 4 \\ 
\frac{1}{\sqrt{1-4/z}} \left( ln \left[ \frac{1 + \sqrt{1-4/z}}{1-\sqrt{1-4/z}} \right] - i \pi \right), & z > 4
\end{array}
\right]
\end{eqnarray}
It is evident from (18) and (20) that this contribution diverges sharply at z = 4 \cite{this}.

Consequently, we replace 
eq. (6) with the following $<\bar qq>$- and $<G^2>$-augmented definition for 
$\Delta F_1$ in order to include the leading nonperturbative quark and gluon condensate 
contributions to the effective Wilson coefficients (eq. (4)):
\begin{eqnarray}
\Delta F_1 (z)&=&-4\int^1_0dxx(1-x)ln\left [1-zx(1-x)\right ] \nonumber\\
&+& \frac{8\pi^2<m_q\bar qq>}{3m_q^4}\left [1-\left (1+\frac{2}{z}\right )\sqrt{1-\frac{4}{z}}\right] 
\nonumber\\
&+& \frac{\pi <\alpha_s G^2>}{6m_q^4}\frac{1}{z^2{(1-\frac{4}{z})}^2}\left [ \frac{48(1-\frac{2}{z})}{z^2} X(z)-4+\frac{16}{z}-\frac{48}{z^2}\right ]
 ,
\end{eqnarray}
where the second term is written in terms of the renormalization group (RG) 
invariant combination $<m_q\bar qq>$, and where the factor $\alpha_s$ in the 
third term is absorbed in the gluon condensate order parameter to achieve approximate RG-invariance.

In the decays $B\to h_1h_2$, where $h_1$ and $h_2$ are light mesons, the characteristic 
range of $k^2$ is argued to be\cite{sw}
\begin{equation}
\frac{m_b^2}{4}\stackrel{<}{\sim}  k^2 \stackrel{<}{\sim} \frac{m_b^2}{2}\;\; .
\end{equation}
For light quarks $\left(z=k^2/m_q^2>>1 \right)$, the nonperturbative terms in eq. (21) can be approximated by
\begin{equation}
\Delta F_1^{\rm nonperturbative}\approx\frac{16\pi^2<m_q\bar qq>}{k^4}-\frac{2\pi <\alpha_s G^2>}{3k^4} \;\; .
\end{equation}
Inserting the phenomenological values $<\bar qq>\approx (-0.25)^3$ GeV$^3$ and 
$<\alpha G^2>\approx 0.045$ GeV$^4$, we observe that the leading nonperturbative corrections 
are around three orders of magnitude smaller than the perturbative 
contribution, and can therefore be safely neglected.  

For the heavier charmed quark, however, 
the situation is more interesting because of the presence of a singularity at $z=4$ 
in the gluon-condensate term:  the range $4 GeV^2 \stackrel{<}{\sim} k^2 \stackrel{<}{\sim} 10 GeV^2$ 
suggested by (22) is quite likely inclusive of $z = 4$, (i.e., of $k^2=4m_c^2$) given the 
present 1.1-1.4 GeV range of values for $m_c$\cite{cea}.  Consequently, there is a genuine possibility
that the gluon-condensate contribution to the penguin amplitude could be comparable to the purely
perturbative correction, depending on which value of $k^2$ one chooses.  This sensitivity
to the momentum transfer $k^2$ is also important for investigating 
direct CP asymmetry in hadronic B decays, as the gluon-condensate contribution to (21) develops
an imaginary part which dominates the strong phase of the effective Wilson 
coefficients as $z \rightarrow 4$ from above.

For a physical interpretation of the above results, we again use the analogy with 
the dileptonic B decays $B\to X_s\ell^+\ell^-$.  These decay modes receive 
a perturbative charmed-quark loop correction (continuum contribution) as well as 
nonperturbative contributions when $c\bar c$ forms intermediate resonances such 
as $J/\psi$, $\psi'$,...\cite{ahm96}.  The latter can be incorporated by 
modelling a Breit-Wigner form for the resonance propagator in the long distance 
amplitude\cite{bw}
\begin{equation}
A^{\rm nonperturbative}\sim\frac{1}{M^2-k^2-iM\Gamma}\;\; ,
\end{equation}
where $k$ is the momentum transfer to dileptons, and where $M$ and $\Gamma$ are 
the mass and total decay width of the intermediate resonance.  The resulting 
dileptonic differential decay rate has peaks at $k^2\sim M^2$, and the nonperturbative 
resonance contribution indeed dominates the branching ratio of this mode.  

One can make a similar charmonium-resonance interpretation for the singularity at 
$k^2 = 4m_c^2$ in the $<G^2>$ contribution to the effective Wilson coefficients of the 
condensate-augmented penguin amplitude, particularily if we identify the $k^2 > 4m_c^2$ kinematic
threshold for the imaginary part with the onset of a "physical" intermediate state.
It must be noted, however, that such a 
resonance couples directly to a single gluon (Fig. 3), necessarily
implying that the $z = 4$ singularity, {\it if indeed a resonance}, must be 
associated with the leading contribution of a weakly bound {\it colour-octet} 
$c \bar{c}$ state\cite{con}.
   
In conclusion, we have calculated the nonperturbative quark and 
gluon condensate contributions to the effective Wilson coefficients 
of the QCD penguin four-fermion operators, supplementing the one 
loop (continuum) $q\bar q$ correction which appears at NLL precision 
with its nonperturbative counterparts.  We have pointed out how incorporation of 
the gluon condensate in the charmed quark loop results in a contribution
which is highly sensitive to momentum transfer $k^2$.  Since $k^2$ is 
not in-and-of itself a physical observable in two body hadronic B decays, 
one cannot exclude the possibility of significant nonperturbative contributions 
to both the magnitude {\it and} the  strong phase of the effective Wilson coefficients 
of the four-Fermi operators $O_3...O_6$.

\section*{Acknowledgements}

MRA would like to thank Emi Kou for useful discussions and comments on quark condensate contribution.  VE is grateful for support from the Natural Sciences and Engineering Research Council of Canada.  MRA acknowledges support from the Science and Technology Agency of Japan. 
\newpage


\newpage

{\center \bf \huge Figure Captions}
\vskip 3.0cm
\noindent
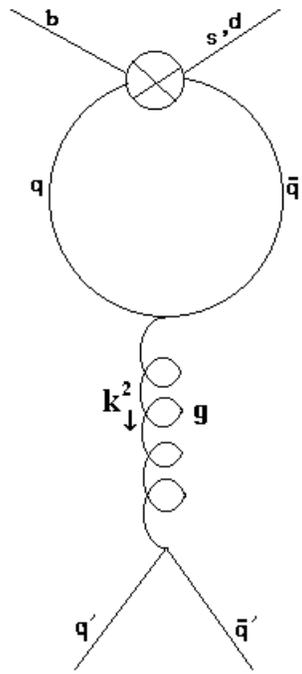
{\bf Figure 1}: Perturbative one loop correction to the effective 
coefficient of the four-fermion operators. \\
\vskip 0.5cm
\noindent
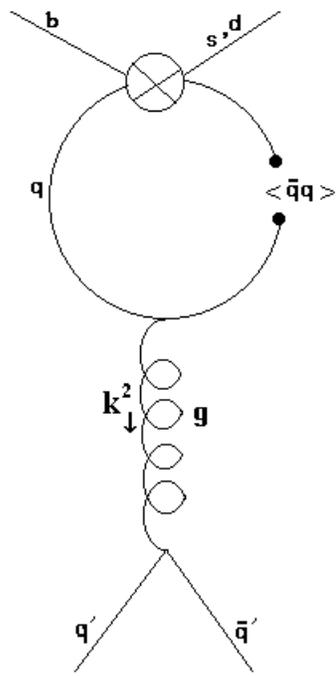
{\bf Figure 2}: Nonperturbative quark condensate contribution to the 
effective coefficient of the four-fermion operators at the one-loop level. \\
\vskip 0.5cm
\noindent
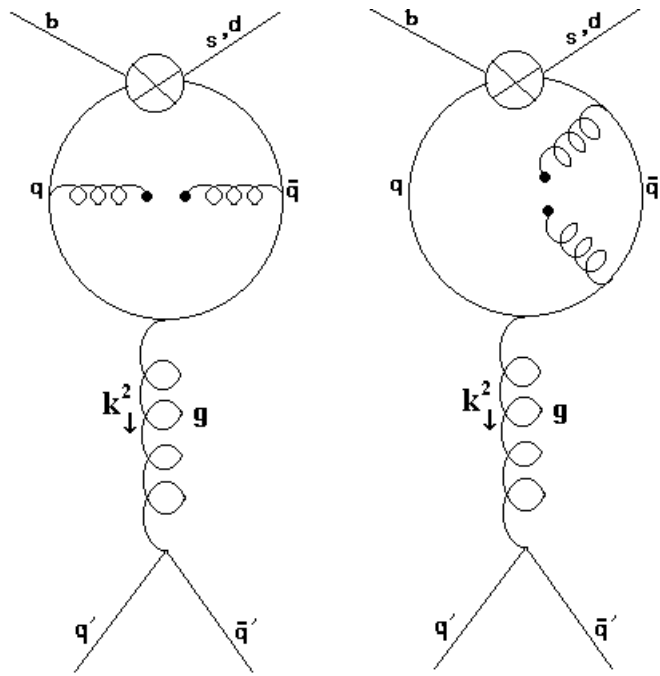
{\bf Figure 3}: Nonperturbative gluon condensate contribution to the 
effective coefficient of the four-fermion operators at the one-loop level. \\
\vskip 0.5cm

\begin{thebibliography}{99}
\bibitem{con} Confinement of such a state presumably entails additional
dynamics.  The idea of colour-nonsinglet intermediate states having some
impact at colour-singlet asymptopia is hardly a new one, though such
states are more typically indentified with quarks and gluons.
\end{thebibliography}
\end{document}